\documentclass[aps,prl,twocolumn,floats]{revtex4}
\usepackage{latexsym}
\usepackage{amsmath}
\usepackage{amssymb}
\usepackage{bm}
\usepackage[dvips]{color}
\usepackage{graphicx}
\usepackage{graphicx}
\usepackage{epsfig}

\newcommand{\hide}[1]{}
\newcommand{\veps}{\varepsilon}

\def\ua{\uparrow}
\def\da{\downarrow}

\def\veps{\varepsilon}

\begin{document}

\title{A Model for Overscreened Kondo Effect in Ultracold Fermi
       Gas}

\author{I. Kuzmenko$^1$, T. Kuzmenko$^1$, Y. Avishai$^{1}$ and
    K. Kikoin$^2$}
\affiliation{\footnotesize
  $^1$Department of Physics, Ben-Gurion University of
  the Negev, Beer-Sheva, Israel \\
  $^2$Raymond and Beverly Sackler Faculty of Exact
  Sciences, School of Physics and Astronomy, Tel Aviv University,
  69978 Tel Aviv, Israel}

\date{\today}

\begin{abstract}
The feasibility of realizing overscreened Kondo effect in
ultra-cold Fermi gas of atoms with spin $s \ge \tfrac{3}{2}$ in
the presence of a localized magnetic impurity atom is proved
realistic. Specifying (as a mere example), to a system of ultra
cold $^{22}$Na Fermi gas  and a trapped $^{197}$Au impurity, the
mechanism of exchange interaction between the Na and Au atoms is
elucidated and the exchange constant is found to be positive
(antiferromagnetic).  The corresponding exchange Hamiltonian is
derived, and the Kondo temperature is estimated at the order of
$1~\mu$K. Within a weak-coupling renormalization group scheme, it
is shown that the coupling renormalizes to the non-Fermi liquid
fixed point. An observable displaying multi-channel
features even in the weak coupling regime is the impurity
magnetization: For ${T}\gg{T}_{K}$ it is {\it  negative}, and then
it increases to become positive with decreasing temperature. 
\end{abstract}

\maketitle

\noindent {\bf{Introduction:}} The experimental discovery of
Bose-Einstein condensation back in 1996 opened the way to study a
myriad of fundamental physical phenomena that were otherwise very
difficult to realize (see Ref. \cite{bloch} for a review). A few
years afterward, fabrication and control a cold gas of {\it
fermionic atoms} has been realized \cite{bourdel,pra04,Jin-06,%
pra06,NuclPhys07,Nature08,Li09,Nature10,Science10}. This
revelation opens the way to study the physics of a gas of fermions
with (half integer) spin $s \ge \tfrac{3}{2}$. The main axis of
the present study relates to the question whether the occurrence
of this new state of matter exposes a new facet of Kondo physics.

Single-channel Kondo effect in cold atom physics has been studied
in Refs. \cite{recati,gupta,chen,ultra-cold-SO3-Kondo,Christoph}.
In Ref. \cite{Lal10}, the possibility of observing multichannel
Kondo effect has been explored  for ultra-cold {\it bosonic atoms}
coupled to an atomic quantum dot, as well as for a system composed
of superconducting nano-wires coupled to a Cooper-pair box.
Recently, non-Fermi liquid behavior has been predicted for Au
monatomic chains containing one Co atom as a magnetic impurity
\cite{prl13}.

In this work we propose a realization of non-Fermi liquid Kondo
effect in cold atoms employing the mechanism leading to
over-screening by large spins. The idea is to localize an atom
with spin $S$ in a gas of cold fermion atoms of spin
$s\ge\frac{3}{2}$ trapped in a combination of harmonic and
periodic potentials. If an exchange interaction
$J{\bf{s}}\cdot{\bf{S}}$ with $J>0$ exists, the underlying Kondo
physics is equivalent to multichannel Kondo effect with large
{\it{effective number of channels}} $N_{s}$\cite{Kim}, that easily
satisfies the Nozi\`{e}res-Blandin inequality $N_{s}>2S$, leading
to over-screening \cite{NB}. Possible candidates for fermionic
atoms are ${}^{22}$Na (electronic spin $\frac{1}{2}$, nuclear spin
$3$ and total atomic spin $s=\frac{5}{2}$), ${}^{40}$K (electronic
spin $\frac{1}{2}$, nuclear spin  $4$ and total atomic spin
$s=\frac{7}{2}$), ${}^{84}$Rb or ${}^{86}$Rb (electronic spin
$\frac{1}{2}$, nuclear spin $2$ and total atomic spin
$s=\frac{3}{2}$). Possible candidate for the impurity is
${}^{197}$Au atom (electronic spin $\frac{1}{2}$, nuclear spin
$\frac{3}{2}$ and total atomic spin $S=1$).

\begin{figure}[htb]
\centering
 \includegraphics[width=50 mm,angle=0]
   {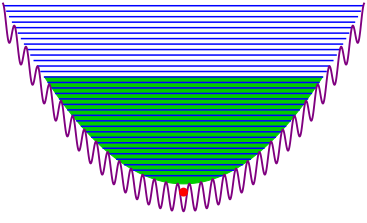}
 \caption{\footnotesize
   Filling of the energy levels in the potential well
   $V_{1D}^{\parallel}(X)$ (\ref{V-opt-1D-def}) by fermionic
   atoms. The filled area denotes the occupied energy levels (blue
   lines), whereas the energy levels in the unfilled area are
   unoccupied. The red circle denotes the impurity. The purple
   curve denotes $V_{1D}^{\parallel}(X)$.}
 \label{Fig-well-filling}
 \vspace{-0.2in}
\end{figure}

\noindent {\bf{Model}}: Typically, ultra-cold fermi gas is stored
in optical dipole traps that rely on the interaction between an
induced dipole moment in an atom and an external electric field
${\bf E}({\bf r},t)$. Such oscillating electric (laser) field
induces an oscillating dipole moment in the atom. Usually, the
trapping potential is formed by three pairs of laser beams of
wavelength $\sim1~\mu$m with the use of an acousto-optic
modulator, creating a time-averaged optical potential
\cite{opt-latt-11,Nature03,arXiv09,prb09,NaturePhys09,%
Greiner-PhD-03}. This technique gives an anisotropic three
dimensional (3D) trap with trapping potential
\begin{eqnarray}
  \label{V1D-anisotropy-def}
  &&
  V_{3D}({\bf{R}}) =
  \sum_{i=1,2}
  V_{1D}^{\parallel}(X_i)+
  V_{1D}^{\perp}(X_3),
\end{eqnarray}
where $X_{1,2,3}$ are Cartesian coordinates of an atom. Each term
on the RHS contains a high-frequency wave which forms the
oscillating potential and a low-frequency wave which forms the
harmonic potential \cite{opt-latt-11,Nature03,arXiv09,prb09,%
NaturePhys09}, (see Fig.~\ref{Fig-well-filling}):
\begin{eqnarray}
  V_{1D}^{p}(X) &=&
  \alpha
  \lim_{T\to\infty}
  \frac{1}{T}
  \int\limits_{0}^{T}dt~
  \big|
      2{\bf{E}}_f(X,t)+
      2{\bf{E}}_p(X,t)
  \big|^2\nonumber\\
  &=&
  V_f\sin^2(k_f X)+
  V_p\big(k_p X\big)^2,
  \label{V-opt-1D-def}
\end{eqnarray}
where $\alpha$ is the electric polarizability of atoms,
$p=\parallel,\perp$, $V_f=2\alpha|{\bf{E}}_f|^2$ and
$V_{p}=2\alpha|{\bf{E}}_{p}|^2$ are the lattice potential depths
which can be controlled by varying the intensities ${\bf{E}}_f$ or
${\bf{E}}_{p}$ of the laser field or the low-frequency waves. The
potential parameters are tuned such that
\begin{equation}
  \label{ineq}
  {V}_{\parallel}k_{\parallel}^2 \ll
  {V}_{\perp}k_{\perp}^2 \ll
  {V}_fk_f^2.
\end{equation}

To be concrete, we henceforth consider fermionic ${}^{22}$Na atoms
(spin $\tfrac{5}{2}$) and ${}^{197}$Au impurities (spin $S=1$).
The potential well is filled with ${}^{22}$Na atoms and with
sparsely distributed ${}^{197}$Au atoms. The shallow well $V_s$
should be deep enough to trap the heavy Au atom but it cannot trap
the Na atoms. As a result, we get an atomic Fermi gas with a low
concentration of magnetic impurities.

\noindent
 \underline{ \bf Atomic Quantum States in the Potential Well}:\\
We consider a neutral atom (nucleus plus core) as a positively
charged rigid  ion (with filled shell) and one electron on the
outer orbital (i.e., 3$s$ orbital in the Na atom or 6$s$ orbital
in the Au atom). The positions of the ion and the outer electron
are respectively specified by vectors ${\bf{R}}$ and ${\bf{r}}$
[Fig.~1 in the supplementary materials (SM)]. In the adiabatic
approximation (which is natural in atomic physics), the wave
function of the atom is a product of the corresponding wave
functions $\Psi({\bf{R}})$ and $\psi({\bf{r}})$ describing the
stationary states of the ion and the outer electron. In order to
find the wave functions and energies of the $^{22}$Na and
$^{197}$Au  ions in the anisotropic 3D potential well, we need to
solve the following Schr\"odinger equation for
$\Psi({\bf{R}})=\Psi^{\rm{Au,Na}}({\bf{R}})$,
\begin{eqnarray}
 -\frac{\hbar^2}{2M}~
  \frac{\partial^2\Psi({\bf{R}})}
       {\partial{\bf{R}}^2}+
  V_{3D}({\bf{R}})
  \Psi({\bf{R}})
  =
  \veps
  \Psi({\bf{R}}).
  \label{eq-Schrodinger-def}
\end{eqnarray}
 Consider first $ \Psi^{\rm{Au}}({\bf{R}})$. When
the corresponding energy level  $\veps_{\rm{imp}}$ is deep enough,
the wave function of the bound state near the potential minimum at
${\bf{R}}={\bf{0}}$ can be approximated within the harmonic
potential picture as,
\begin{eqnarray}
  \Psi^{\rm{Au}}({\bf{R}}) &=&
  \frac{1}{\big(\pi a_f^2\big)^{\frac{3}{2}}}~
  \exp
  \bigg(-
       \frac{R^2}{2a_f^2}
  \bigg),
  \label{wave-fun-Au}
\end{eqnarray}
where
\begin{eqnarray*}
  k_f a_f =
  \sqrt{\frac{\hbar\omega_f}{V_f}},
  \ \ \
  \omega_f =
  \sqrt{\frac{2V_f k_f^2}{M_{\rm{Au}}}},
  \ \ \
  \veps_{\rm{imp}} =
  \frac{3\hbar\omega_f}{2}.
  \label{imp-level}
\end{eqnarray*}
Second, consider the wave function of the (much lighter) $^{22}$Na
ions, for which  the shallow potential wells are not deep enough
to form bound states. For studying the Kondo effect  we need to
focus on quantum states at energies $\veps$ within the deep well
close to $\epsilon_F$, that is, $\veps\gg{V_f}$. In that case we
can neglect the ``fast'' potential relief $V_f\sin^2(k_f x)$, and
the solution of Eq. (\ref{eq-Schrodinger-def}) becomes,
\begin{subequations}
\begin{eqnarray}
  \Psi_{\nu m \ell}^{\rm{Na}}({\bf{R}}) &=&
  \Phi_{\nu m}(R)~
  F_{\ell}(Z)~
  e^{im\phi},
  \label{Psi-3D-res}
\end{eqnarray}
where $R$, $\phi$ and $Z$ are cylindrical coordinates. Denoting
$\rho\equiv R/a_\parallel$, the radial wave function $\Phi_{\nu
m}(R)$ is,
\begin{eqnarray}
  \Phi_{\nu m}(R) =
\frac{1}{a_{\parallel}\sqrt{\pi}}~
  \sqrt{\frac{\nu!}{(\nu+|m|)!}}
      \rho^{|m|}
  L_{\nu}^{(|m|)}\big(\rho^2\big)
  e^{\mbox{-}\frac{\rho^2}{2}},
  \label{wave-fun-radial-harm}
\end{eqnarray}
where $L_{\nu}^{(|m|)}$ is the generalized Laguerre polynomial,
$\nu=0,1,2,\ldots$ and $m=0,\pm1,\pm2,\ldots$, and
\begin{eqnarray}
  k_{\parallel} a_{\parallel}=
  \sqrt{\frac{\hbar\omega_{\parallel}}{V_{\parallel}}},
  \ \ \ \ \
  \omega_{\parallel} =
  \sqrt{\frac{2V_{\parallel} k_{\parallel}^2}{M_{\rm{Na}}}}.
  \label{a-s-omega-s-def}
\end{eqnarray}
Denoting $\zeta \equiv Z/a_\perp$ the motion along $Z$ is
described by
\begin{eqnarray}
  F_{\ell}(Z) &=&
 \frac{1}{\big(\pi a_{\perp}^2\big)^{\frac{1}{4}}}
  \frac{1}{\sqrt{2^{\ell}~\ell!}}
  H_{\ell}(\zeta)
  e^{-\zeta^2/2},
  \label{wave-fun-Z-harm}
\end{eqnarray}
where $H_{\ell}$ is the Hermite polynomial, $\ell$ is the harmonic
quantum number, $\ell=0,1,2,\ldots$.
\begin{eqnarray}
  &&
  k_{\perp} a_{\perp}=
  \sqrt{\frac{\hbar\omega_{\perp}}{V_{\perp}}},
  \ \ \ \ \
  \omega_{\perp} =
  \sqrt{\frac{2V_{\perp} k_{\perp}^2}{M_{\rm{Na}}}}.
  \label{a-perp-omega-perp-def}
\end{eqnarray}
  \label{subeqs-wave-fun-slow}
\end{subequations}
The corresponding energy levels depend on two quantum number,
$n=2\nu+|m|$ and $\ell$,
\begin{eqnarray}
  \veps_{n\ell} =
  \hbar\omega_{\parallel}
  \Big(
      n+
      1
  \Big)+
  \hbar\omega_{\perp}
  \Big(
      \ell+
      \frac{1}{2}
  \Big).
  \label{energy-levels}
\end{eqnarray}
When $\omega_{\parallel}$ and $\omega_{\perp}$ are incommensurate,
the degeneracy of the level $(n,\ell)$ is $(n+1)(2s+1)$. The
inequalities (\ref{ineq}) imply
$\omega_{\parallel}\ll\omega_{\perp}\ll\omega_f$. Restricting the
Fermi energy $\epsilon_F$ as,
$\frac{\hbar\omega_{\perp}}{2}<\epsilon_F<%
\frac{3\hbar\omega_{\perp}}{2}$, the quantum states with $\ell>0$
are frozen at $T\ll\hbar\omega_{\perp}$, hence the Fermi gas is
virtually 2D.

The potential well is filled by $^{22}$Na atoms (fermions) and one
impurity atom $^{197}$Au. The latter occupies the lowest-energy
level (\ref{imp-level}) of the potential well and its wave
function is well concentrated around the point
${\bf{R}}={\bf{0}}$.  Hence, regarding it as a localized impurity
is justified. \\
\noindent {\bf  Na-Au exchange interaction}: When the distance
between a $^{22}$Na atom and the $^{197}$Au impurity is of the
order  of $R_0$ (the atomic size), there is an exchange
interaction between their open electronic $s$ shells
\cite{Andreev73}.  It includes a direct exchange term of strength
$J_d$ (due to antisymmetrization of the electronic wave functions
where electrons do not hop between atoms), and an indirect
exchange term of strength $J_p$ (due to contribution from polar
states, where electrons can hope between atoms). Unlike the case
of hydrogen molecule where the direct part dominates, here both of
them  should be considered since their orders of magnitude are
found to be comparable.
Evaluating the exchange interaction between Na and Au atoms
involves four wave functions: Two of them,
$\Psi^{\rm{Au}}({\bf{R}})$ [Eq. (\ref{wave-fun-Au})] and
$\Psi^{\rm{Na}}_{{\nu}{m}{0}}({\bf{R}})$ [Eq. (\ref{Psi-3D-res})]
pertain to the corresponding atoms as being structureless
particles in the optical potential (\ref{V-opt-1D-def}).  The
other two $\psi_3^{\rm{Na}}({\bf{r}})$ and
$\psi_6^{\rm{Au}}({\bf{r}})$ pertain to electronic wave functions
of the 3s orbital in Na and 6s orbital in Au,
\begin{eqnarray}
  J_{\nu\nu',m} =
  -\frac{2}{(2s+1)(2S+1)}~
  \int\limits_{R_{12}>R_0}
  d^{3}{\mathbf{R}}_{1}
  d^{3}{\mathbf{R}}_{2}~
  V(R_{12})
  \nonumber \\ \times
  \Big|
      \Psi^{\mathrm{Au}}({\mathbf{R}}_{1})
  \Big|^2
  \Big(
      \Psi^{\mathrm{Na}}_{\nu m 0}({\mathbf{R}}_{2})
  \Big)^{*}
  \Psi^{\mathrm{Na}}_{\nu' m 0}({\mathbf{R}}_{2}), ~~~~~ ~~~
  \label{exchange-def}
\end{eqnarray}
where ${\mathbf{R}}_{1}$ or ${\mathbf{R}}_{2}$ is the position of
the ion of Au or Na, $R_{12}=|{\mathbf{R}}_{1}-{\mathbf{R}}_{2}|$.
We assume here that $R_{12}>R_0$, since Coulomb repulsion between
the electron clouds of the Na and Au atoms prevents the atoms from
approaching closer than
${R}_{0}\approx{r}_{\mathrm{Na}}+{r}_{\mathrm{Au}}$ (that is
approximated by the sum of the corresponding atomic radii).
$V(R)=V_{\mathrm{d}}(R)+\frac{W_{\mathrm{p}}^{2}(R)}{U}$, where
$V_{\mathrm{d}}(R)$ is a direct exchange interaction,
$W_{\mathrm{p}}(R)$ is the hybridization term and $U$ is a Coulomb
blockade. Explicitly they are (see Ref. \cite{Davydov-QM} and the
SM for details),
\begin{eqnarray*}
  V_{\mathrm{d}}(R_{12}) &=&
  \int
  d^3{\bf{r}}_1
  d^3{\bf{r}}_2~
  \psi_6^{\rm{Au}}({\mathbf{r}}_{1}-{\mathbf{R}}_1)
  \psi_3^{\rm{Na}}({\mathbf{r}}_{2}-{\mathbf{R}}_2)
  \nonumber \\ && \times
  \psi_3^{\rm{Na}}({\mathbf{r}}_{1}-{\mathbf{R}}_2)
  \psi_6^{\rm{Au}}({\mathbf{r}}_{2}-{\mathbf{R}}_1)
  \nonumber \\ && \times
  \bigg\{
       \frac{e^2}{{r}_{12}}+
       V_{12}(R_{12})+
       \sum_{i=1}^{2}
       V_{i}(|{\mathbf{r}}_{i}-{\bf{R}}_{i}|)
  \bigg\},
  \\
  W_{\rm{p}}(R_{12}) &=&
  \frac{1}{2}
  \int{d^3{\bf{r}}}~
  \sum_{i=1}^{2}
  V_{i}(|{\bf{r}}-{\bf{R}}_{i}|)
  \nonumber \\ && \times
  \psi^{\rm{Au}}_6({\mathbf{r}}-{\mathbf{R}}_{1})
  \psi^{\rm{Na}}_3({\mathbf{r}}-{\mathbf{R}}_{2}),
  \\
  \frac{1}{U} &=&
  \frac{1}{U_{\rm{Na}}+\veps_{\rm{Na}}-\veps_{\rm{Au}}}+
  \frac{1}{U_{\rm{Au}}+\veps_{\rm{Au}}-\veps_{\rm{Na}}}.
\end{eqnarray*}
Here ${\bf{r}}_1$ or ${\bf{r}}_2$ is the position of electron and
${r}_{12}=|{\bf{r}}_1-{\bf{r}}_2|$. $V_{1}(r)$ or $V_{2}(r)$
describes the electron-ion interaction for Au or Na, and
$V_{12}(R)$ is the interaction between ions.
$\veps_{\rm{Na}}=-5.14$~eV and $\veps_{\rm{Au}}=-9.23$~eV are
single-electron energies of the sodium and gold atoms,
$U_{\rm{Na}}=5.69$~eV and $U_{\rm{Au}}=11.54$~eV are the Coulomb
interaction preventing two-electron occupation of the outer
orbitals of atoms.

The electronic wave functions decrease rapidly when the distance
between the atoms exceeds the atomic radius, so that the exchange
interaction may be approximated by a point-like interaction.
Moreover, the wave function $\Psi^{\rm{Au}}({\bf{R}})$,
Eq.(\ref{wave-fun-Au}), has its maximum at ${\bf{R}}={\bf{0}}$ and
it vanishes for ${R}\gg{a_f}$. The wave function
$\Psi^{\rm{Na}}_{{\nu}{m}{0}}({\bf{R}})$, Eq.(\ref{Psi-3D-res}),
varies slowly on the distance scale of $a_f$. Then
$|\Psi^{\rm{Au}}({\bf{R}})|^2$ can be approximated by the
delta-function. Within this approximation, we get the following
estimate of the exchange constant, $J_{\nu\nu',m}=J\delta_{m0}$,
where
\begin{eqnarray}
  J &=&
  \frac{1}
       {(2s+1)(2S+1)}~
  \frac{\hbar^2(u_{\mathrm{d}}+u_{\mathrm{p}})}
       {\sqrt{\pi}~
        M_{\mathrm{Na}}~
        a_{\parallel}^2~
        a_{\perp}},
  \label{exchange-Na-Au-res}
\end{eqnarray}
where $u_{\mathrm{d}}$ and $u_{\mathrm{p}}$ are scattering lengths
of the direct and indirect exchange interactions. Expressions for
$u_{\mathrm{d}}$ and $u_{\mathrm{p}}$ in terms of the potentials
of the direct and indirect exchanges as well as the explicit form
for the exchange potentials are standard and can be found in
textbooks [see {\it e.g.} Refs.
\cite{Sitenko,Landau-Lifshitz-3,Davydov-QM} and Eqs. (12) and (13)
in the Supplementary Material]. Numerical estimations yields
$u_{\mathrm{d}}\approx1.24~\mu$m and $u_{\mathrm{p}}=1.49~\mu$m.
In Refs. \cite{Landau-Lifshitz-3,Davydov-QM} it was shown that
$J_{\mathrm{d}}>0$. Since $J_{\mathrm{p}}>0$ (always), the total
exchange interaction is anti-ferromagnetic.

\noindent {\bf{Kondo Hamiltonian and the Kondo temperature}}: Eq.
(\ref{exchange-Na-Au-res}) indicates that, due to centrifugal
barrier, only  atoms in s-states interact with the impurity.
Omitting the quantum numbers $\ell=m=0$ for brevity, we write the
Hamiltonian of the system as $H=H_0 + H_K$, where
\begin{equation}
  H_0 =
  \sum_{\nu\mu}
  \veps_{2\nu}~
  c_{\nu\mu}^{\dag}
  c_{\nu\mu},
  \ \ \ \ \
  H_K =
  J~
  \big(
      {\bf{S}}
      \cdot
      {\bf{s}}
  \big).
  \label{Hex-Kondo-def}
\end{equation}
Here  $c_{\nu\mu}$ or $c_{\nu\mu}^{\dag}$ is the annihilation or
creation operator of a sodium atom in the state with the principal
quantum number $2\nu$ and $\ell=m=0$,
$\veps_{2\nu}\equiv\veps_{2\nu0}$ is given by Eq.
(\ref{energy-levels}), $\mu$ denotes atomic spin projection,
${\bf{S}}$ is the impurity spin and
 ${\bf{s}} =
  \sum_{\nu\nu',\mu\mu'}
  c_{\nu\mu}^{\dag}
  {\mathbf{t}}_{\mu\mu'}
  c_{\nu'\mu'},$ where
$\hat{\mathbf{t}}$ is the vector of the spin-$s$ matrices.

The density of states (DOS) for the Hamiltonian $H_0$ is
\begin{eqnarray}
  \rho(\epsilon) &=&
  \sum_{\nu=0}^{\infty}
  \delta(\epsilon-\veps_{2\nu})
  =
  \frac{\vartheta(\epsilon)}
       {2\hbar\omega_{\parallel}},
  \label{DOS-def}
\end{eqnarray}
where $\vartheta(\epsilon)$ is the Heaviside theta function.

Within poor-man scaling formalism for multy-channel Kondo effect,
the dimensionless coupling $j=J\rho(\epsilon_F)$ satisfies the
following scaling equation  \cite{NB},
\begin{eqnarray}
  \frac{\partial{j}(D)}
       {\partial\ln{D}}
  &=&
  -j^2(D)+
  N_sj^3(D),
  \label{scal-eq}
\end{eqnarray}
with $N_s=\frac{2}{3}s(s+1)(2s+1)$ being an {\it effective number
of channels} \cite{Kim}. Initially, the bandwidth is
${D_0}\ge{D}\gg{T}$ and the initial value of $j(D)$ is,
\begin{eqnarray*}
  &&
  j(D_0)\equiv
  j_0=
  \frac{1}{(2s+1)(2S+1)}~
  \frac{u_d+u_p}{2\sqrt{\pi}~a_{\perp}},
\end{eqnarray*}
where  $a_{\perp}$ is given by Eq.(\ref{a-perp-omega-perp-def}).

The solution of Eq.(\ref{scal-eq}) is,
\begin{eqnarray}
  \ln
  \bigg(
       \frac{D_0}{D}
  \bigg)
  =
  \frac{1}{j_0}-
  \frac{1}{j}+
  N_s
  \ln
  \frac{j(1-N_s j_0)}
       {j_0(1-N_s j)}.
  \label{j(T)}
\end{eqnarray}
When $D \to 0$, $j(D)$ renormalizes to the  weak coupling fixed
point $j^{*}=1/N_s$. When $|j(D)-j^{*}|\ll{j^{*}}$, the solution
for $j(D)$ reduces asymptotically to,
\begin{eqnarray}
  \frac{j^{*}-j(D)}{j^{*}} =
  \frac{j^{*}-j_0}{j_0}~
  \bigg(
       \frac{D~T^{*}}{D_0T_{K}}
  \bigg)^{j^{*}},
  \label{j(T)-asymptot}
\end{eqnarray}
 see Ref. \cite{Hewson-book}, where
\begin{eqnarray}
  T^{*} =
  D_0
  \exp\bigg(-\frac{1}{j^{*}}\bigg),
  \ \ \
  T_{K} =
  D_0
  \exp\bigg(-\frac{1}{j_0}\bigg).
  \label{TK}
\end{eqnarray}
The scaling equation (\ref{scal-eq}) accounts for the evolution of
$j(D)$ only when the atomic spin $s\ge\frac{3}{2}$.
\begin{figure}[htb]
\centering
\includegraphics[width=60 mm,angle=0]
   {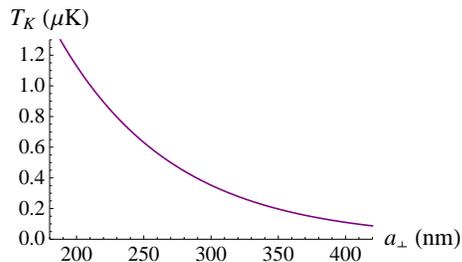}
 \caption{\footnotesize
   $T_K$, Eq.~ (\ref{TK}) as a function of $a_{\perp}$ for
   $D_0\approx{V_f}=1$~neV.}
   \vspace{-0.2 in}
 \label{Fig-TK}
\end{figure}
The Kondo temperature (\ref{TK}) as a function of $a_{\perp}$ is
shown in Fig.~\ref{Fig-TK} for $D_0\approx{V_f}=1$~neV. It is seen
that $T_K$ changes from $100$~nK to $1.1~\mu$K for
$a_{\perp}=200\div400$~nm, so that the ratio
$\frac{T_K}{D_0}\approx0.01\div0.1$ is really small, whereas
${T}\sim{T}_{K}$ may be experimentally reachable. Indeed, more
than a decade ago, $^{40}$P atoms were cooled to a temperature of
50~nK \cite{Nature03}, later the ${}^{133}$Cs atoms were cooled to
temperature 40~nK \cite{NaturePhys09,arXiv09}.

\noindent {\bf{Impurity magnetization}}: Having elaborated upon
the theory, we are now in a position to carry out perturbation
calculations of experimental observables.  It is sometimes argued
that the interesting physics in the over-screens Kondo effect is
exposed only in the strong coupling regime. Here we show that
peculiar behaviour emerges also in the weak coupling regime. The
reason is that the weak coupling fixed point $j^*$ is small, and
in most cases, the initial value of $j(D_0) > j^*$. As the
temperature $T$ is reduced toward $T_K$, $j(D)$ {\it decreases}
toward $j^*$ and as a results, some physical observables display
an unusual dependence on temperature. Consider for example the
impurity magnetization ${\bf{M}}_{\rm{imp}}(T)=M_{\rm{imp}}(T)
\hat {\bf B}$ in response to an external magnetic field ${\bf B}$.
Experimentally it requires immersing a small concentration $n_i$
of impurity atoms in the gas of fermionic atoms. Within third
order perturbation theory, we have,
\begin{eqnarray}
 && M_{\rm{imp}} =
  \frac{B\chi_0T^{*}}{T}~
  \Bigg\{
       \frac{S}{S+1}-
       \frac{S N_s}{s+1}~
       \bigg[
            j+
  \nonumber \\
  && +
            j^2
            \Big(
                1-j N_s
            \Big)
            \ln\bigg(\frac{D_0}{T}\bigg)
       \bigg]
  \Bigg\},
  \ \
  \chi_0=
  \frac{g_e^2 \mu_B^2 n_i}{12T^{*}},
  \label{chi-third-order}
\end{eqnarray}
where $g_e\approx2$ is the electronic spin g-factor and $\mu_B$ is
the Bohr magneton.

Due to the logarithmic terms, which, strictly speaking, are not
small either, the terms proportional to $j^2$ and $j^3$ are not
small as compared with $j$.  Hence, expansion up to the third
order in $j$ is inadequate. Instead, we derive an expression for
the impurity related magnetization in the leading logarithmic
approximation using the RG equations (\ref{scal-eq}). The
condition imposing invariance of the magnetization under ``poor
man's scaling'' transformation has the form \cite{Hewson-book},
\begin{equation}
  \label{dchi-dlogD}
  \frac{\partial}{\partial D}
  \bigg[
       j+
       j^2
       \Big(
           1-j N_s
       \Big)
       \ln\bigg(\frac{D}{T}\bigg)
  \bigg]=0.
\end{equation}
Eq.~(\ref{dchi-dlogD}) yields the scaling equation
(\ref{scal-eq}). The renormalization procedure should proceed
until the bandwidth $D$ is reduced to the temperature $T$. The
expression for the impurity related magnetization then becomes,
\begin{equation}
  M_{\rm{imp}} =
  \frac{B\chi_0 T^{*}}{T}
  X(T),
  \ \
  X(T)=
  \frac{S}{S+1}-
  \frac{S}{s+1}~
  \frac{j(T)}{j^{*}}.
  \label{chi-scal}
\end{equation}

The function $X(T)$ consists of two terms. The first one describes
the Zeeman interaction of the impurity with the external magnetic
field and results in the Curie's law. The second one corresponds
to the exchange interaction of the impurity with atoms (the atomic
magnetization is parallel to the external magnetic field). When
the exchange interaction of the impurity is stronger than the
Zeeman interaction, the function $X(T)$ is negative and the
impurity magnetization is anti-parallel to the external magnetic
field. This occurs when $j(T)$ exceeds some critical value,
$j_{c}=j^{*}\frac{s+1}{S+1}$.

\begin{figure}[htb]
\centering
 \includegraphics[width=60 mm,angle=0]
   {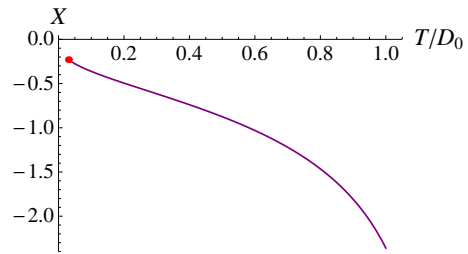}
 \caption{\footnotesize
   The function $X(T)$, Eq. (\ref{chi-scal}), as a function
   of temperature for $j_0=0.286$. The red dot corresponds to $T=T_K$.}
 \label{Fig-chi}
 \vspace{-0.1in}
\end{figure}

Fig.~\ref{Fig-chi} illustrates $X(T)$ for $j_0=0.286$. It is seen
that at high temperatures when $j(T)>j_{c}$, $X$ is negative, and
the impurity magnetization is anti-parallel to the external
magnetic field.  With reducing the temperature, the effective
coupling $j(T)$ reduces as well. At temperature $T_c$ satisfying
$j(T_c)=j_c$, $X(T)$ changes sign from negative for $T>T_c$ to
positive for $T<T_c$. For the given parameter values,
 $T_c \ll T_K$ and,
strictly speaking, cannot be estimated within the framework of the
poor man's scaling technique.

\noindent {\bf Conclusions:} The non-Fermi liquid Kondo effect can
be accessed within the realm of cold atom physics. Exchange
Hamiltonian is derived and scaling equations are solved for an
ultra-cold gas of alkali atoms [such as ${}^{22}$Na] with
${}^{197}$Au  impurity. The dimensionless coupling $j$ is not
extremely small even though the coupling $J$, Eq.
(\ref{exchange-Na-Au-res}), is small. Such
over-screened Kondo effect by fermions of large spin may be
exposed even in the weak coupling regime through the temperature
dependence of the impurity magnetization. \\
\ \\
\noindent {\bf Acknowledgement:} We would like to thank N. Andrei,
O. Parcollet, Y. Castin and C. Salomon for important  discussions
and numerous suggestions during the early stages of this research.
This work is supported by grant 400/2012 of the Israeli Science
Foundations (ISF).

\vspace{-0.2in}

\newpage

\onecolumngrid

\appendix

\section{Supplementary Material}
Here we expand upon the derivation of exchange constants between
$^{22}$Na and $^{197}$Au that are required to arrive at the Kondo
Hamiltonian (10) of the main text. First we elucidate the direct
exchange and then the indirect one. As it turn out, both of them
are positive for realistic inter-atomic distance $R_{12} $ and they
are of the same order of magnitude.

\subsection{Direct Exchange Contribution}

\begin{figure}[htb]
\centering
\includegraphics[width=55 mm,angle=0]
   {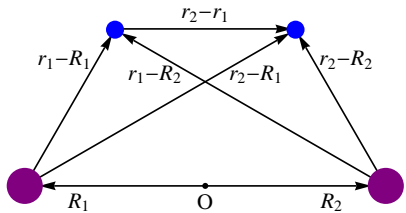}
 \caption{\footnotesize
   Two atoms. Position of electron of the
   first or second atom is ${\bf{r}}_1$ or ${\bf{r}}_2$,
   the radius vector between the nuclei is ${\bf{R}}$.}
 \label{Fig-atoms}
\end{figure}

Let atoms of gold and sodium be at positions ${\mathbf{R}}_{1}$
and ${\mathbf{R}}_{2}$, with the distance between them
$R_{12}=|{\mathbf{R}}_{1}-{\mathbf{R}}_{2}|$. Then the direct
exchange interaction $V_{\mathrm{d}}(R_{12})$ between the atoms
is (see Ref.[29]),
\begin{eqnarray}
  V_{\rm{d}}(R_{12}) &=&
  \int
  d^3{\bf{r}}_1
  d^3{\bf{r}}_2~
  \psi_6^{\rm{Au}}({\mathbf{r}}_{1}-{\mathbf{R}}_{1})
  \psi_3^{\rm{Na}}({\mathbf{r}}_{2}-{\mathbf{R}}_{2})
  \psi_3^{\rm{Na}}({\mathbf{r}}_{1}-{\mathbf{R}}_{2})
  \psi_6^{\rm{Au}}({\mathbf{r}}_{2}-{\mathbf{R}}_{1})
  \times \nonumber \\ && \times
  \bigg\{
       \frac{e^2}{{r}_{12}}+
       V_{12}(R_{12})+
       V_{1}(|{\mathbf{r}}_{1}-{\bf{R}}_{1}|)+
       V_{2}(|{\mathbf{r}}_{2}-{\mathbf{R}}_{2}|)
  \bigg\}.
  \label{ex-atom}
\end{eqnarray}
Here ${\bf{r}}_1$ or ${\bf{r}}_2$ is the position of electron, see
Fig.~\ref{Fig-atoms} and ${r}_{12}=|{\bf{r}}_1-{\bf{r}}_2|$.
We assume here that $R_{12}>R_0$, since Coulomb repulsion between
the electron clouds of the Na and Au atoms prevents the atoms from
approaching closer than
${R}_{0}\approx{r}_{\mathrm{Na}}+{r}_{\mathrm{Au}}$ (that is
approximated by the sum of the corresponding atomic radii).  In
Eq.~(\ref{ex-atom}), $V_{1}(r)$ or $V_{2}(r)$
describes the electron-ion interaction for Au or Na, and
$V_{12}(R)$ is the interaction between ions. When the
inter-atomic distance exceeds $R_0$, we can write,
$$
  V_{1}(r)
  \approx
  V_{2}(r)
  \approx
  -\frac{e^2}{r},
  \ \ \ \ \
  V_{12}(R)
  \approx
  \frac{e^2}{R}.
$$

\begin{figure}[htb]
\centering
\includegraphics[width=65 mm,angle=0]
   {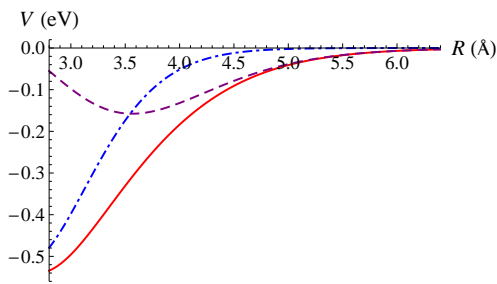}
 \caption{\footnotesize
   Direct exchange interaction $V_{\mathrm{d}}(R)$
   [Eq.~(\ref{ex-atom}), dashed purple curve],
   indirect exchange interaction $V_{\mathrm{p}}(R)$
   [Eq.~(\ref{ex-indir-atom}), dashed and dotted blue curve]
   and the total exchange interaction $V(R)$
   [Eq. (\ref{V=Vdir+Vpol}), solid red curve]
   as functions of the distances $R$ between the nuclei.}
   \vspace{-0.1in}
 \label{Fig-exchange}
\end{figure}

The function $V_{\mathrm{d}}(R)$ calculated numerically for the
hydrogen-like electronic wave functions
$\psi_3^{\rm{Na}}({\bf{r}})$ and $\psi_6^{\rm{Au}}({\bf{r}})$ is
shown in Fig. \ref{Fig-exchange}, dashed purple curve. It is negative
for any $R>R_0$ [where $R_0\approx3.3$~{\AA} for the atoms of
sodium and gold], so that the exchange interaction is anti-ferromagnetic.

\subsection{Indirect Exchange Contribution}

Indirect exchange interaction between the atoms Na and Au separated by
distance $R_{12}$ is
\begin{eqnarray}
  V_{\mathrm{p}}(R_{12}) &=&
  -\frac{W_{\rm{p}}^2(R_{12})}{U},
  \label{ex-indir-atom}
\end{eqnarray}
where
$$
  \frac{1}{U} ~=~
  \frac{1}{U_{\rm{Na}}+\veps_{\rm{Na}}-\veps_{\rm{Au}}}+
  \frac{1}{U_{\rm{Au}}+\veps_{\rm{Au}}-\veps_{\rm{Na}}},
$$
$\veps_{\rm{Na}}=-5.14$~eV and $\veps_{\rm{Au}}=-9.23$~eV are
single-electron energies of the sodium and gold atoms,
$U_{\rm{Na}}=5.69$~eV and $U_{\rm{Au}}=11.54$~eV are the Coulomb
interaction preventing two-electron occupation of the outer
orbitals of atoms.

The hybridization term $W_{\mathrm{p}}(R)$ in Eq.
(\ref{ex-indir-atom}) is given explicitly as,
\begin{eqnarray}
  W_{\rm{p}}(R) &=&
  \frac{1}{2}
  \int{d^3{\bf{r}}}~
  \Big\{
      V_{1}(|{\bf{r}}|)+
      V_{2}(|{\bf{r}}-{\bf{R}}|)
  \Big\}~
  \psi^{\rm{Au}}_6({\bf{r}})
  \psi^{\rm{Na}}_3({\bf{r}}-{\bf{R}}).
  \label{tunnel-rate-def}
\end{eqnarray}

The function $V_{\mathrm{p}}(R)$ calculated numerically for the
hydrogen-like electronic wave functions $\psi_3^{\rm{Na}}({\bf{r}})$
and $\psi_6^{\rm{Au}}({\bf{r}})$ is shown in Fig. \ref{Fig-exchange},
dashed and dotted blue curve. It is  is always negative, so that the indirect exchange
interaction is anti-ferromagnetic.

\subsection{Projecting the Exchange Interaction onto the States
with a Given Total Spin}

The exchange interaction Hamiltonian can be written as,
\begin{eqnarray}
  H_{\mathrm{ex}}(R_{12}) &=&
  -2V(R_{12})~
  \big(
      {\mathbf{s}}_{1}
      \cdot
      {\mathbf{s}}_{2}
  \big),
  \label{H-exch-spin-electron}
\end{eqnarray}
where
\begin{eqnarray}
  V(R) &=&
  V_{\mathrm{d}}(R)+
  V_{\mathrm{p}}(R).
  \label{V=Vdir+Vpol}
\end{eqnarray}
$V(R)$ calculated numerically for the hydrogen-like electronic wave
functions $\psi_3^{\rm{Na}}({\bf{r}})$ and $\psi_6^{\rm{Au}}({\bf{r}})$
is shown in Fig. \ref{Fig-exchange}, solid red curve. It is
seen that $V(R)<0$, so that the coupling is anti-ferromagnetic.
${\mathbf{s}}_{1}$ or ${\mathbf{s}}_{2}$ is spin operator for
the outer s-electron of the gold or sodium atom,
$$
  {\mathbf{s}}_{j} =
  \frac{1}{2}
  \sum_{\sigma\sigma'}
  d_{j\sigma}^{\dag}
  {\boldsymbol\tau}_{\sigma\sigma'}
  d_{j\sigma'},
$$
where $\hat{\boldsymbol\tau}$ is a vector of the Pauli matrices,
$d_{j\sigma}$ or $d_{j\sigma}^{\dag}$ is the annihilation or
creation operator of electron with spin $\sigma=\ua,\da$.

Atom of ${}^{197}$Au has the nuclear spin $s_{\rm{Au}}=\frac{3}{2}$,
so the quantum state of  an atom, $|\sigma,\mu\rangle$, is described
by projection of the nuclear spin $\mu$ on the axis $z$ and electronic
spin $\sigma$. Anti-ferromagnetic hyperfine interaction couples nuclear
and electron spins in total atomic spin $s=s_{\mathrm{Au}}-\frac{1}{2}=1$.
The wave function $|S,m\rangle$ of the state with the total spin $S$
and the projection of the spin $m$ on the axis $z$ is,
\begin{eqnarray}
  \big|
      S,m
  \big\rangle
  &=&
  \sqrt{\frac{S+1-m}{2(S+1)}}~
  \bigg|
       \ua,m-\frac{1}{2}
  \bigg\rangle-
  \sqrt{\frac{S+1+m}{2(S+1)}}~
  \bigg|
       \da,m+\frac{1}{2}
  \bigg\rangle.
  \label{WF-spin-el-nucl}
\end{eqnarray}

Projecting out the electronic spin operator ${\mathbf{s}}_{1}$
onto the quantum states (\ref{WF-spin-el-nucl}), we get
\begin{eqnarray}
  {\mathbf{s}}_{1} &\to&
  \sum_{mm'}
  \big|
      S,m
  \big\rangle
  \big\langle
      S,m
  \big|
      {\mathbf{s}}_{1}
  \big|
      S,m'
  \big\rangle
  \big\langle
      S,m'
  \big|
  ~=~
  \frac{\mathbf{S}}{2S+1},
  \ \ \ \ \
  {\mathbf{S}} ~=~
  \sum_{mm'}
  {\mathbf{T}}_{mm'}~
  X^{mm'}_{\mathrm{Au}},
  \label{spin-projecting-Au}
\end{eqnarray}
where $\hat{\mathbf{T}}$ is a vector of the spin-$s$ matrices,
\begin{eqnarray*}
  X^{mm'}_{\mathrm{Au}} &=&
  \big|
      S,m
  \big\rangle
  \big\langle
      S,m'
  \big|.
\end{eqnarray*}

Similarly, the nuclear spin of $^{22}$Na is $s_{\mathrm{Na}}=3$,
the total atomic spin is $s=3-\frac{1}{2}=\frac{5}{2}$. The wave function
$|s,m\rangle$ of the quantum state with the total spin $s$ and
projection $m$ of the spin on the $z$-axis is given by Eq.(\ref{WF-spin-el-nucl})
with ${S}\to{s}$. Then projecting out the electronic spin operator ${\mathbf{s}}_{2}$
onto the quantum states (\ref{WF-spin-el-nucl}), we get
\begin{eqnarray}
  {\mathbf{s}}_{2} &\to&
  \sum_{mm'}
  \big|
      s,m
  \big\rangle
  \big\langle
      s,m
  \big|
      {\mathbf{s}}_{1}
  \big|
      s,m'
  \big\rangle
  \big\langle
      s,m'
  \big|
  ~=~
  \frac{\mathbf{s}}{2s+1},
  \ \ \ \ \
  {\mathbf{s}} ~=~
  \sum_{mm'}
  {\mathbf{t}}_{mm'}~
  X^{mm'}_{\mathrm{Na}},
  \label{spin-projecting-Na}
\end{eqnarray}
where $\hat{\mathbf{t}}$ is a vector of the spin-$s$ matrices,
\begin{eqnarray*}
  X^{mm'}_{\mathrm{Na}} &=&
  \big|
      s,m
  \big\rangle
  \big\langle
      s,m'
  \big|.
\end{eqnarray*}

Finally, the exchange Hamiltonian (\ref{H-exch-spin-electron}) takes
the form,
\begin{eqnarray}
  H_{\mathrm{ex}}(R_{12}) &=&
  -\frac{2V(R_{12})}{(2s+1)(2S+1)}~
  \big(
      {\mathbf{s}}
      \cdot
      {\mathbf{S}}
  \big).
  \label{H-exch-spin-total}
\end{eqnarray}

\subsection{Derivation of the Coupling $J$}

Atoms of sodium and gold place in the external potential given by Eq. (1)
of the main text. The wave function $\Psi^{\mathrm{Au}}({\mathbf{R}})$ of
the atom of gold is given by Eq. (5) of the main text, whereas the wave
functions $\Psi^{\mathrm{Na}}_{\nu m 0}({\mathbf{R}})$ of the atoms of
sodium are given by Eq.(6a) of the main text. Then the coupling is,
\begin{eqnarray}
  J_{\nu\nu',m} =
  -\frac{2}{(2s+1)(2S+1)}~
  \int\limits_{R_{12}\ge{R_0}}
  d^3{\bf{R}}_1
  d^3{\bf{R}}_2~
  V(R_{12})~
  \Big|
      \Psi^{\rm{Au}}
      \big(
          {\bf{R}}_1
      \big)
  \Big|^2~
  \Big(
      \Psi^{\rm{Na}}_{\nu m 0}
      \big(
          {\bf{R}}_2
      \big)
  \Big)^{*}~
  \Psi^{\rm{Na}}_{\nu'm 0}
  \big(
      {\bf{R}}_2
  \big),
  \label{ex-Na-Au-def}
\end{eqnarray}
where $V(R)$ is given by Eq.(\ref{V=Vdir+Vpol}). The sign of $J_{\nu\nu',m}$
is chosen in such a way that positive coupling strength corresponds to
anti-ferromagnetic interaction. The integration on the RHS
of Eq.(\ref{ex-Na-Au-def}) is restricted by the condition
$R_{12}>R_0$, since Coulomb repulsion between the
electron clouds of the Na and Au atoms prevents the atoms from
approaching closer than $R_0$.

The function $V(R)$ is negative for any $R>R_0$ (see Fig.\ref{Fig-exchange}),
so that the exchange
interaction is anti-ferromagnetic. $|V(R)|$ has its maximum at some value
${R}\sim{R}_{0}$ and vanishes when ${R}\gg{R_0}$. The atomic wave
functions $\Psi^{\rm{Au}}({\bf{R}})$ and $\Psi^{\rm{Na}}_{{\nu}{m}{0}}({\bf{R}})$
change slowly at a range of $R_0$. Therefore, the following approximations
are justified:
(1) changing the limits of integration on the RHS of Eq.(\ref{ex-Na-Au-def})
 from $R_0\le{R}_{12}<\infty$ to $0\le{R}_{12}<\infty$ and
 (2) approximating $V(R)$ by a delta function,
\begin{eqnarray}
  &&
  V({R}_{12})
  \approx
  -V_0~
  \delta({\bf{R}}_{12}),
  \nonumber \\
  &&
  V_0 =
  -4\pi~
  \int\limits_{R_0}^{\infty}
  V(R)
  R^2dR =
  \frac{2\pi\hbar^2}{M_{\mathrm{Na}}}~
  \big(
      u_{\mathrm{d}}+
      u_{\mathrm{p}}
  \big),
  \label{exchange-delta-approx}
\end{eqnarray}
where $u_{\mathrm{d}}$ and $u_{\mathrm{p}}$ are scattering lengths
of the direct and indirect exchange interactions,
\begin{eqnarray}
  u_{\mathrm{d}} &=&
  -\frac{2M_{\mathrm{Na}}}{\hbar^2}
  \int\limits_{R_0}^{\infty}
  V_{\mathrm{d}}(R)
  R^2dR,
  \label{scattering-length-direct}
  \\
  u_{\mathrm{p}} &=&
  -\frac{2M_{\mathrm{Na}}}{\hbar^2}
  \int\limits_{R_0}^{\infty}
  V_{\mathrm{p}}(R)
  R^2dR.
  \label{scattering-length-indir}
\end{eqnarray}
Numerical estimate with hydrogen-like electronic wave functions
$\psi_3^{\rm{Na}}({\bf{r}})$ and $\psi_6^{\rm{Au}}({\bf{r}})$,
${V}_{\mathrm{Na}}(r)\approx{V}_{\mathrm{Au}}(r)\approx-\frac{e^2}{r}$
and $V_{\mathrm{ion}}(R)\approx\frac{e^2}{R}$
yields $u_{\mathrm{d}}\approx1.24~\mu$m and
$u_{\mathrm{p}}=1.49~\mu$m.

The Au wave function $\Psi^{\rm{Au}}({\bf{R}})$ [Eq.(5) of the main text]
has its maximum at ${\bf{R}}={\bf{0}}$ and it vanishes for ${R}\gg{a_f}$.
The wave function $\Psi^{\rm{Na}}_{{\nu}{m}{0}}({\bf{R}})$ [Eq.(6a) of
the main text] changes slowly on the distance scale of $a_f$.
Then the function $|\Psi^{\rm{Au}}({\bf{R}})|^2$ in Eq.(\ref{ex-Na-Au-def})
 can be approximated by the delta-function,
\begin{eqnarray}
  |\Psi^{\rm{Au}}({\bf{R}})|^2
  &\approx&
  \delta({\bf{R}}).
  \label{Psi-imp-delta-approx}
\end{eqnarray}
Substituting Eqs. (\ref{exchange-delta-approx}) and
(\ref{Psi-imp-delta-approx}) into Eq. (\ref{ex-Na-Au-def}),
we get the following estimate of the exchange constant,
\begin{eqnarray*}
  &&
  J_{\nu\nu',m} ~\approx~
  J~
  \delta_{m0},
\end{eqnarray*}
where $J$ is given by Eq. (9) of the main text.

\end{document}